\documentclass[preprint,showpacs,preprintnumbers,amsmath,amssymb,superscriptaddress]{revtex4}

\usepackage{graphicx}
\usepackage{dcolumn}
\usepackage{bm}

\begin{document}


\title{Transport scattering time probed through rf admittance of a graphene capacitor}

\author{E. Pallecchi}
\email{Emiliano.Pallecchi@lpa.ens.fr}
\affiliation{Laboratoire Pierre Aigrain, Ecole Normale Sup\'erieure, CNRS (UMR 8551),
Universit\'e P. et M. Curie, Universit\'e D. Diderot,
24, rue Lhomond, 75231 Paris Cedex 05, France}
\author{A.C. Betz}
\affiliation{Laboratoire Pierre Aigrain, Ecole Normale Sup\'erieure, CNRS (UMR 8551),
Universit\'e P. et M. Curie, Universit\'e D. Diderot,
24, rue Lhomond, 75231 Paris Cedex 05, France}
\author{J. Chaste}
\affiliation{Laboratoire Pierre Aigrain, Ecole Normale Sup\'erieure, CNRS (UMR 8551),
Universit\'e P. et M. Curie, Universit\'e D. Diderot,
24, rue Lhomond, 75231 Paris Cedex 05, France}
\author{G. F\`eve}
\affiliation{Laboratoire Pierre Aigrain, Ecole Normale Sup\'erieure, CNRS (UMR 8551),
Universit\'e P. et M. Curie, Universit\'e D. Diderot,
24, rue Lhomond, 75231 Paris Cedex 05, France}
\author{B. Huard}
\affiliation{Laboratoire Pierre Aigrain, Ecole Normale Sup\'erieure, CNRS (UMR 8551),
Universit\'e P. et M. Curie, Universit\'e D. Diderot,
24, rue Lhomond, 75231 Paris Cedex 05, France}
\author{T. Kontos}
\affiliation{Laboratoire Pierre Aigrain, Ecole Normale Sup\'erieure, CNRS (UMR 8551),
Universit\'e P. et M. Curie, Universit\'e D. Diderot,
24, rue Lhomond, 75231 Paris Cedex 05, France}
\author{J.-M. Berroir}
\affiliation{Laboratoire Pierre Aigrain, Ecole Normale Sup\'erieure, CNRS (UMR 8551),
Universit\'e P. et M. Curie, Universit\'e D. Diderot,
24, rue Lhomond, 75231 Paris Cedex 05, France}
\author{B. Pla\c{c}ais}
\affiliation{Laboratoire Pierre Aigrain, Ecole Normale Sup\'erieure, CNRS (UMR 8551),
Universit\'e P. et M. Curie, Universit\'e D. Diderot,
24, rue Lhomond, 75231 Paris Cedex 05, France}

\date{\today}

\begin{abstract}

We have investigated electron dynamics in top gated graphene by measuring the gate admittance of a diffusive graphene capacitor in a broad frequency range as a function of carrier density. The density of states, conductivity and diffusion constant are deduced from the low frequency gate capacitance, its charging time and their ratio. The admittance evolves from an RC-like to a skin-effect response at GHz frequency with a crossover given by the Thouless energy. The scattering time is found to be independent of energy in the $0-200\;\mathrm{meV}$ investigated range at room temperature. This is consistent with a random mass model for Dirac Fermions.

\end{abstract}

\maketitle

\section{Introduction}

Electron scattering in graphene is still a debated matter (see \cite{CastroRMP2009,DasSarmaRMP2010,PeresRMP2010} and references therein).  The situation is complicated in this 2-dimensional crystal by the strong influences of the substrate, the surface contamination, the effects of static distortions and phonons \cite{Mozorov2008PRL,Dean2010Nnano,NovoselovScience2004,Tan2007PRL,Chen2008Nnano,Chen2008Nphys,Chen2009PRL}. By contrast investigation   technics are rather limited; the main diagnosis  relies  on the temperature and carrier density ($n_c$) dependencies of the conductivity.
The former is weak for graphene on substrates and attributed to phonons  \cite{Chen2008Nphys,Zhu2009PRB,Efetov2010arXiv}.  The later comes from the dependence of the scattering time $\tau(k_F)$ on wave number which can be varied in a broad range due to efficient electrostatic gate doping. We have listed in TABLE-\ref{table1} the most predicted behaviors for the $\tau(k_F)$ and the corresponding $\sigma(n_c)$ laws \cite{Shon1998JPSJ,Aleiner2006PRL,Ziegler2006PRL,Nomura2007PRL,Ostrovsky2006PRB,Katsnelson2008PhilTrans,Huang2008PRB}. They can be classified in two categories corresponding to a nearly linear density dependence of the conductivity (or $\tau\sim k_F$), or a constant conductivity ($\tau\sim k_F^{-1}$).  The former is exemplified by the charge impurity mechanism and the latter by the local impurity model. Intermediate situations can be explained by an admixture of both mechanisms giving rise to a sublinear density dependence of conductivity. Alternatively sublinearity could  be accounted for by a single mechanism, like the  Dirac-mass disorder associated  with a random lifting of sublattice degeneracy \cite{Cheianov2010EPL}. Indeed, according to  Ref.\cite{Ziegler2006PRL}, a random Dirac-mass mechanism should give $\tau=Const.$.
It translates into a $\sigma\propto \sqrt{n_c}$  dependence of the bulk conductivity, which should not be confused with the $G\propto\sqrt{n_c}$ dependence of a two terminal conductance in ballistic graphene \cite{Bolotin2008SSC,Du2008Nnano}.

Experimentally a linear dependence of conductivity is reported at low temperature with a tendency to sublinearity at high density. It is well explained by a Boltzmann conductivity  $\sigma^{-1}(n_c)=(n_c e\mu+\sigma_0)^{-1}+\rho_s$ \cite{Dean2010Nnano}, with a mobility $\mu= ev_F\tau/\hbar k_F$,  a saturation at $\rho_s^{-1}$ and a conductivity minimum $\sigma_0$. The weak effect of dielectric environment \cite{Ponomarenko2009PRL} and  complementary measurements of the quantum scattering time \cite{Monteverde2010PRL} suggest an interpretation in terms of resonant scattering.

\begin{table}[htbp]
  \begin{tabular}{|l|l|l|l|} \tableline
   \emph{ mechanisms} &   \emph{scattering time} & \emph{conductivity }& \emph{Ref.}\\ \tableline
local impurity &   $\tau\sim 1/k_F$ & $ \sigma\sim Const$ & \cite{Shon1998JPSJ}\\ \tableline
local impurity &   $\tau\sim\ln{k_F}/k_F$ & $\sigma\sim\ln{n_c}$  & \cite{Aleiner2006PRL}\\ \tableline
random Dirac-mass  &   $\tau\sim Const$ & $\sigma\sim\sqrt{n_c}$ &  \cite{Ziegler2006PRL} \\ \tableline
charged impurity &   $\tau\sim k_F$ & $\sigma\sim n_c$ & \cite{Nomura2007PRL}\\ \tableline
resonnant scattering &   $\tau\sim k_F\ln^2(k_F)$ & $\sigma\sim n_c\ln^2n_c$ & \cite{Ostrovsky2006PRB}\\ \tableline
ripples &   $\tau\sim k_F^{(2H-1)}$ & $\sigma\sim n_c^{H}$ & \cite{Katsnelson2008PhilTrans}\\ \tableline
acoustic phonons &   $\tau\sim k_F^2$ & $\sigma\sim n_c^{3/2}$ & \cite{Huang2008PRB}\\ \tableline
  \end{tabular}
  \caption{Main scattering mechanisms proposed for graphene classified according to the carrier density dependence :  Fermi wave vector dependence of the scattering time $\tau(k_F)$  and carrier density dependence  of conductivity $\sigma(n_c)$. For the ripples model, $H\sim1$  is the exponent of the correlation function for the lattice distortions.  The expression for acoustic phonons scattering  corresponds to a low temperature limit.}
  \label{table1}\end{table}

In the present work we consider the case of top gated graphene at room temperature. We investigate scattering using the rf gate admittance.  From the in-phase and out-of-phase responses we obtain the thermal averages of the density of states $\rho(E_F)$ and the bulk conductivity $\sigma(E_F)$ as function of energy $E_F=\hbar k_F v_F$ ($v_F\simeq10^6\;\mathrm{m/s}$).  We rely on the Einstein relation $\sigma=e^2\rho {\cal D}$  to deduce the diffusion constant ${\cal D}(E_F)$  and the scattering time, $\tau=2{\cal D}/v_F^2$. We shall focus on the hole-doped regime, $E_F=-(0$-$200)\;\mathrm{meV}$, where we find  that $\tau$ is energy independent. Finally we discuss the possible origin of this observation.

\section{Experimental principles}

In an rf transport experiment, $\rho(E)$ can be obtained from the quantum capacitance $c_Q(E_F)=e^2\int{\rho(E)(-\partial f/\partial E)dE}$, where $E_F$ stands now (and below) for the chemical potential. $c_Q$ is the thermal average of $\rho(E)$ and corresponds to the electronic compressibility, $\chi\equiv\partial n/\partial E_F=c_Q/e^2$ where $n=\int{\rho(E)f(E)dE}$. The finite compressibility is responsible for a chemical contribution $\Delta E_F$ in addition to the electrostatic one in the electronic charging energy.  For a capacitor charge $q=-e\Delta n$ and bias $U$, $\Delta E_F=\Delta n/\chi=-eq/c_Q$ adds the electrostatic term, $-eq/c_{geo}$, in the total energy $-eU=-eq/c_{g}$. Here $c_{geo}$ and $c_{g}$ are the geometrical and  total gate capacitance (per unit area) so that one finally has  $c_g(E_F)^{-1}=c_{geo}^{-1}+c_Q(E_F)^{-1}$. The quantum capacitance term is generally negligible in  metals ($c_Q\rightarrow\infty$) or in back gated conductors ($c_{geo}\rightarrow0$).

 Using the above definition, one can rewrite the Einstein relation as $\sigma=c_q{\cal D}$ \cite{DasSarmaRMP2010}.  $c_Q(E_F)$ can be separated from the constant geometrical capacitance in graphene thanks to its energy dependence \cite{Fang2007APL,Chen2008IEEE,Young2010arXiv,Xia2009Nnano,Droscher2010APL,Ponomorenko2010preprint}. At finite temperature the theoretical expression is  \cite{Fang2007APL}
   \begin{equation}
c_Q= \frac{2 e^2 k_BT}{\pi(\hbar v_F)^2}\times \ln[2+2\cosh(\frac{E_F}{k_BT})] \label{Cq}\; .
\end{equation}
Using nanometers thick gate oxides, $c_{geo}\sim10\;\mathrm{fF/\mu m^2}$ becomes comparable to the quantum capacitance minimum, $c_Q(0)=4 e^2 k_BT\ln[2]/\pi(\hbar v_F)^2\simeq 10\;\mathrm{fF/\mu m^2}$ at neutrality and room temperature.

Top gated graphene offers the possibility to manipulate ac gate currents for dynamical characterization \cite{Chaste2008NanoL} which is also useful for technological applications \cite{Lin2010Science,Schwierz2010Nnano}. It has however the drawback of a  non-linear $n_c(V_{g})$ relation between gate charge and voltage due to the quantum capacitance contribution that complicates the standard analysis.  At $T=0$ in pure graphene one has $n_c=n_{gate}-n_Q[(1+2n_{gate}/n_Q)^{0.5}-1]$ where $n_{gate}=c_{geo}V_{g}/e$ (see e.g. Ref.\cite{DasSarmaRMP2010}) and  $n_Q=\frac{\pi}{2}(c_{geo}\hbar v_F)^2/e^4\simeq 2\times10^{11}\;\mathrm{cm^{-2}}$  ($c_{geo}\simeq10\;\mathrm{fF/\mu m^2}$) entails strong deviations from linear $\sigma(n_{gate})$ behavior. In experiment we shall express the measured quantities directly as function of the chemical potential $E_F$  using the relation
\begin{equation}
E_F(V_{g})= e(V_{g}-V_{np})-\int_{V_{np}}^{V_{g}}{c_Q/(c_{geo}+c_Q)dV}\label{E_F} ,
\end{equation}
where $V_{np}$ is the gate voltage at the neutrality point.
Top gating  gives in principle the possibility to characterize charged impurity scattering by observing a cutt-off in the Thomas-Fermi screening at wave vectors $q_{TF}$ ($q_{TF}\gtrsim 1.5 k_F$ for graphene between SiO$_2$ and AlOx) smaller than a $q_{gate}=1/2t_{ox}$ \cite{Akkermans2004}. However, with $t_{ox}\simeq8\;\mathrm{nm}$, $q_{gate}\simeq6\times10^5\;\mathrm{cm^{-1}}$ still remains at the lower end of our investigated wave number range $k_F=0$--$30\times10^5\;\mathrm{cm^{-1}}$, where thermal and impurity effects are prominent.

\begin{figure}
\centerline {\includegraphics[scale=0.7]{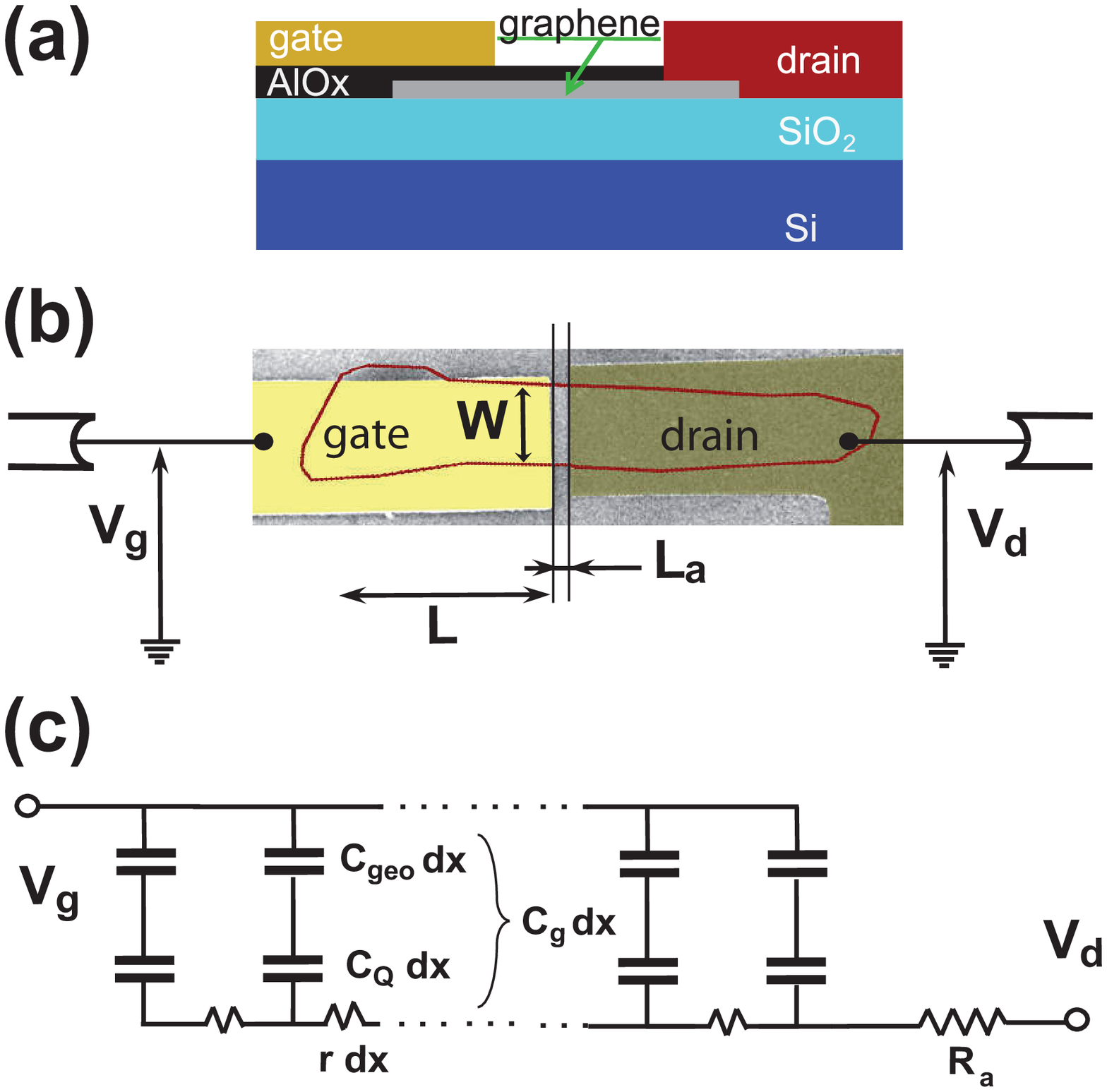}}
\caption{Graphene capacitor made of a monolayer graphene strip coupled to a metallic gate through a thin  oxide (thickness $t_{ox}\simeq 8\;\mathrm{nm}$, permittivity $\kappa\sim7$).  (a) sketch of the layout. (b) SEM picture of sample E9-Zc showing the palladium drain, the gold top gate and the outline of the graphene flake.   (c)   1-dimensional lumped element description of the graphene capacitor, used for data analysis. It includes a distributed resistance $r=\sigma^{-1}/W$ and capacitance $c=c_gW$ where
 $c_g=c_ {geo}c_q/(c_ {geo}+c_q)$ is  the gate capacitance and $c_{geo}$ (resp. $c_q$) its geometrical (resp. quantum) contributions. $R_a$  is a series access resistance.  }
 \label{schema}
\end{figure}

The capacitor geometry of our experiment is described in Fig.\ref{schema}. We have measured several graphene samples, both in the two-terminal capacitor and three-terminal transistor geometries, obtained by the exfoliation method on a thermally oxidized silicon substrate with a resistivity $\rho \geq 20 \;\mathrm{k\Omega~cm}$. AFM inspection, performed prior to gate deposition, shows a significant roughness of the silicon oxide, which contrasts with the relative smoothness of the area covered by graphene.  Data presented here refer to two representative capacitor samples :   samples E9-Zc and C7-F are  flakes of dimensions $L\times W\simeq3\times1\;\mathrm{\mu m^2}$ (see Fig.\ref{schema}) and $L\times W=2\times0.6\;\mathrm{\mu m^2}$.   The use of high resistivity silicon is important to minimize the spurious conductance through the substrate. This parasitic contribution can be very high at microwave frequencies, and effectively shunt the contribution of a small graphene sample. We use palladium for the drain electrode to minimize contact resistance and a gate oxide obtained by multistep oxidation of thin aluminum, for a nominal oxide thickness of about $t_{ox}\simeq8\;\mathrm{nm}$, and finally gold-gate deposition. The distance $L_a$ between the drain and the gate is reduced to $\simeq 200\; \mathrm{nm}$, \textit{i.e.} $\sim6\%$ of the gate length, to minimize the contribution of the access resistance $R_a$. With a permittivity $\kappa\simeq 7$ for AlOx we estimate $c_{geo} \simeq 8\;\mathrm{fF/\mu m^2}$.
The device is inserted in a coplanar wave guide used for RF  characterization \cite{Chaste2008NanoL}.

The samples are measured in an RF probe station at room-temperature. Bias tees are used to control the DC gate-drain voltage $V_{g}$. The RF scattering parameters are measured  with a network analyzer and used to calculate the gate-drain admittance $Y(\omega)$ \cite{Pozar2005}.  The background contribution, which corresponds to a parallel gate-drain capacitive coupling, $C_0\simeq1.8\;\mathrm{fF}$, is measured in an identical but dummy structure and subtracted.  Linear response conditions are secured by probing the device well below the thermal noise floor with an excitation voltage $V_{rf}\sim 1-10\;\mathrm{mV}$.
 We have measured admittance spectra $Y(\omega)$ up to $10\;\mathrm{GHz}$ in a gate voltage range $V_{g}=0$--$1\;\mathrm{V}$ corresponding to average electron  and hole densities $n_c =0$-$2\times  10^{12}\;\mathrm{cm^{-2}}$.  For a quantitative analysis, we shall concentrate in the following on the hole region where the contact resistance due to chemical hole doping of the palladium can be neglected \cite{Lee2008Nnano,Huard2008PRB}.

\section{Experimental results and analysis}

\begin{figure}
\centerline{\includegraphics[scale=0.6]{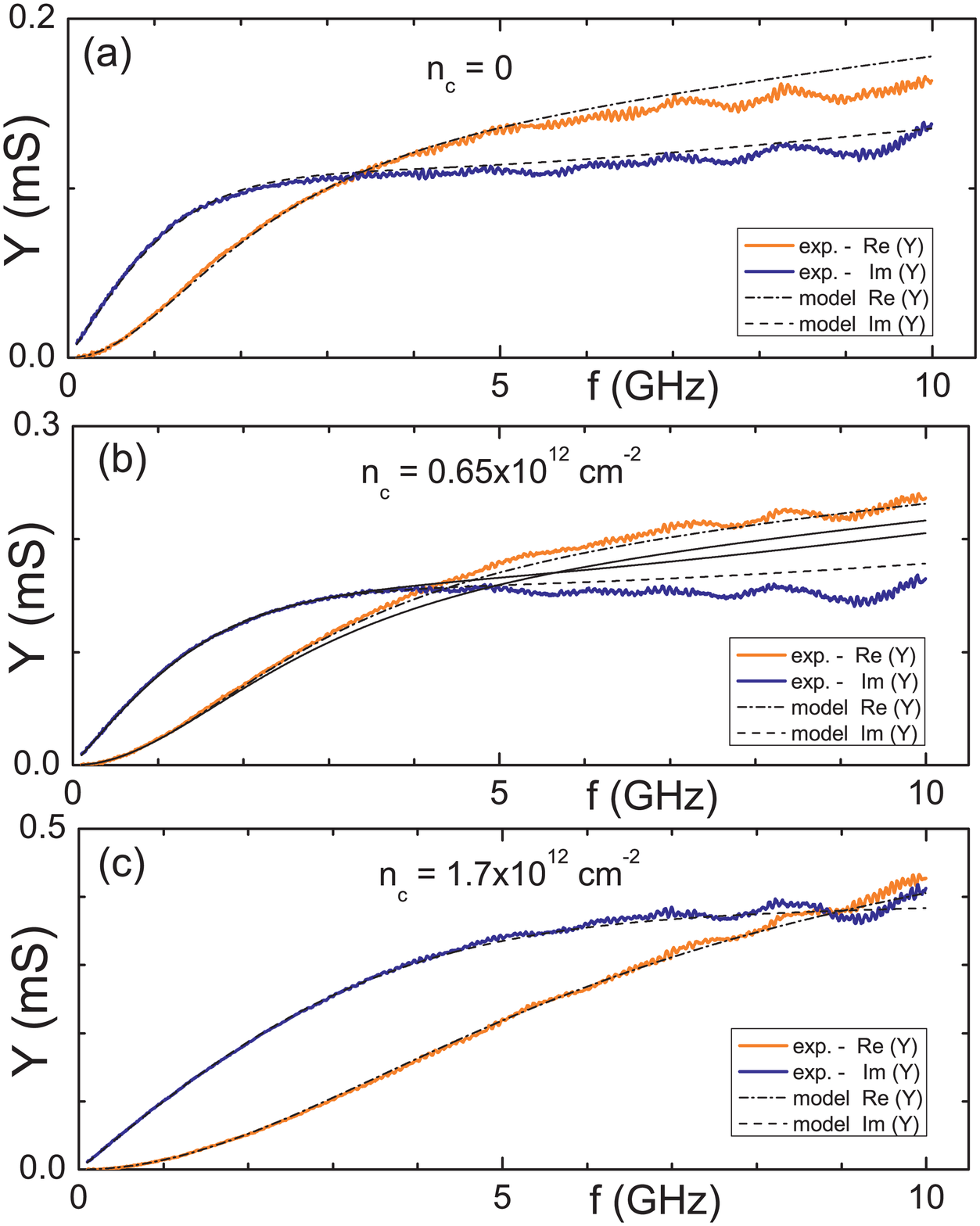}}
\caption{Admittance spectra of a diffusive graphene capacitor (sample E9-Zc) for three gate voltages $V_{g}=(0.68,0.48,0.18)\;\mathrm{V}$ for the (a), (b) and (c) panels. The signature of skin effect (\textit{i.e.} $Re(Y)\simeq Im(Y)\propto\sqrt{f}$) is seen at neutrality ($V_{g}=0.68\;\mathrm{V}$) in panel (a) due to a larger resistivity. At all investigated densities, the admittance spectra are  accurately fitted using the 1-dimensional strip-line model and a small access resistance $R_a=0.15 R_g$.  We find respectively $(C_g,R_g)=(12.5, 4.18);(14.0,2.64);(16.0,1.28)\;\mathrm{(fF,k\Omega)}$. For comparison we have added in panel (b) the fit obtained with $R_a=0$ (dashed lines).      } \label{admittance}
\end{figure}

Fig.\ref{admittance} shows typical admittance spectra of sample E9-Zc for three representative gate voltages corresponding to charge neutrality (a), intermediate (b) and large (c) hole concentrations. The purely capacitive response ($Re(Y)\simeq0$, $Im(Y)\simeq \jmath \omega C_g $) is observed at low frequency.   Strong deviations from this limit are seen above $1\;\mathrm{GHz}$ which are due to finite charge relaxation resistance. Solid lines are fits to the data using a 1-dimensional distributed-rc model (Fig.\ref{schema}-b), with  linear capacitance and resistance  $c=c_gW$ and $r=(\sigma W)^{-1}$. The model gives
\begin{equation}
Y(\omega) = \jmath \omega cL \times\left[\frac{\tanh(\jmath qL)}{\jmath qL}\right] \label{Yg}\; ,
\end{equation}
where $q=(-\jmath rc\omega)^{0.5}$ ($rc=c_g/\sigma$) is the wave number of the probing rf field in the capacitor. We stress that the 1-dimensional character of the  rf probing field penetrating the capacitor should not be confused with the 2-dimensional electronic diffusion probed by the rf field. This 1D field distribution is a powerful simplification in the analysis of the capacitor response as compared to the more complicated 2-dimensional current distribution in a high frequency drain-source conductivity measurement.

The low frequency development,
\begin{equation}
Y(\omega) \simeq \jmath \omega C_g + R_{g} (\omega C_g)^2 + \emph{o}(\omega^3)\label{low_freq}\; ,
\end{equation}
gives the equivalent R-C circuit description which corresponds to an homogeneous charge distribution with a  capacitance, $C_g=c_gLW$, and a resistance, $R_{g}=\sigma^{-1}L/3W$. Eq.(\ref{low_freq}) will be used for the experimental determination of $c_g$ and $\sigma$.

The high frequency limit of Eq.(\ref{Yg}) corresponds to an inhomogeneous charge distribution along the capacitor length corresponding to a penetration over a depth, $\delta=(2\sigma/c_g\omega)^{0.5}\ll L$, from the capacitor drain edge. The asymptotic admittance,
\begin{equation}
Y=\frac{(1+\jmath)}{2}\sqrt{2W^2\sigma c_g\omega }\label{high_freq}\;
\end{equation}
 has a constant phase $\arg(Y)=\pi/4$  and a $\sqrt{\omega}$-dependent modulus. It is markedly different from that of an RC circuit with a frequency independent charging resistance, such as a contact resistance $R_c$ which gives $\Re{(Y)}\rightarrow R_c^{-1}$ and  $\Im{(Y)}\rightarrow 0$. The observation of the asymptotic skin effect regime given by Eq.(\ref{high_freq}) will be taken as a proof that the measured conductivity is the bulk contribution equivalent to that measured in a four terminal dc experiment.

The crossing of real and imaginary parts  at  $\omega_c\simeq \frac{\pi^2}{2}\sigma/c_gL^2$ defines the cutoff frequency of the capacitor. It can be expressed as the sum  $\omega_c= \frac{\pi^2}{2}\sigma/c_{geo}L^2+\frac{\pi^2}{2}{\cal D}/L^2$. The first term is the cutoff of a classical capacitor,  the second is a mesoscopic correction due to the finite density of state. The correction dominates at charge neutrality and low temperature in top gated devices ($\sigma/c_{geo}{\cal D}=e^2\rho(E_F)/c_{geo}\rightarrow 0$). One can note that in these conditions, $\omega_c$  becomes a  rather direct measurement of the diffusion constant, which is a further example of a mesoscopic effect  where a microscopic property (the Thouless energy $\hbar {\cal D}/L^2$) shows up in a macroscopic measurement (the cutoff frequency of a capacitor).

\begin{figure}
\includegraphics[scale=1]{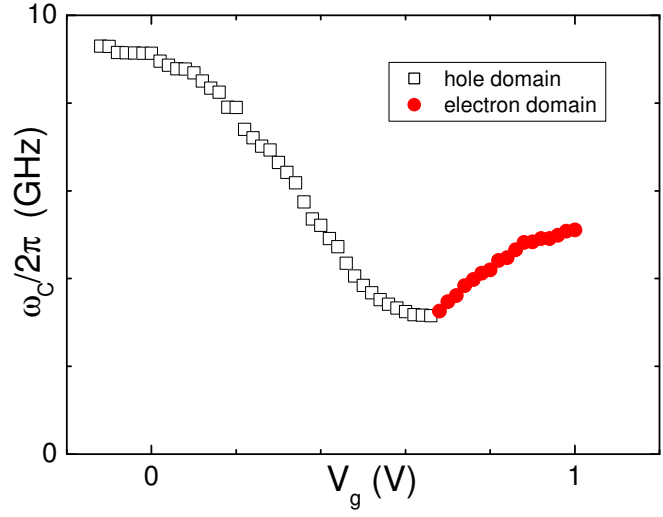}
\caption{Gate voltage dependence of the  cutoff frequency $\omega_c/2\pi$ of sample E9-Zc estimated by the crossing of the real and imaginary parts of the admittance spectrum $Y(\omega)$. The hole region shows a trend to saturation at neutrality ($V_{np}\simeq0.67\;\mathrm{V}$ from which we deduce an (under-) estimate of the diffusion constant ${\cal D}(V_{np})\lesssim L^2\omega_c/5\simeq400\;\mathrm{cm^2/s}$.   } \label{cutoff}
\end{figure}

As shown in Fig.\ref{admittance}, both regimes are observed in experiment with a cutoff that increases with carrier density and levels off at $3\;\mathrm{GHz}$ at neutrality (see Fig.\ref{cutoff}). In particular, the full spectrum of Eq.(\ref{Yg}) can be verified at low hole doping demonstrating the prominent contribution of diffusion in charge relaxation.
In the analysis we have also included the effect of the access region as a serial resistance $R_a$. We approximate $R_a \simeq \sigma^{-1} L_a / W \sim 0.15 R_g$.
The inclusion of $R_a$ results in a small correction in the investigated frequency range, but becomes prominent at very high frequency with an asymptotic limit  $Y\rightarrow R_a^{-1}$ and $\arg(Y)\rightarrow0$ for $\delta\ll L_a$. In order to quantify the effect of $R_a$ we have added in Fig.\ref{admittance}~(b) (dashed line) the spectrum obtained taking $R_a = 0$.
Sample C7-F shows similar spectra with,  however, a higher cutoff due to larger diffusion constant and smaller gate length.

\begin{figure}
\includegraphics[scale=1]{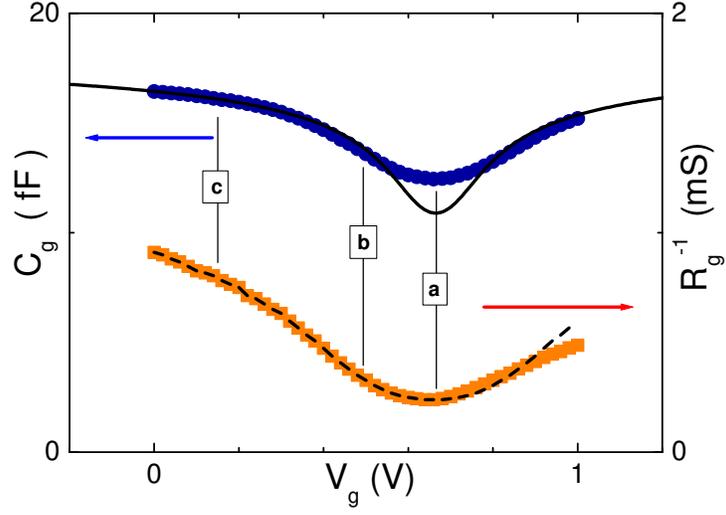}
\caption{Gate capacitance $C_g$ and charge relaxation conductance $R_{g}^{-1}$ of sample E9-Zc deduced from fits of the admittance spectra with Eq.(\ref{Yg}). Labels refer to the admittance spectra in Fig.\ref{admittance} for hole densities $n_c=(0,0.65,1.7)\times 10^{12}\;\mathrm{cm^{-2}}$.  The solid line is a theoretical fit to the data using Eq.(\ref{Cq}) with $C_{geo}=19.2\;\mathrm{fF}$.   The dashed line (lower curve) is a guide for the eye representing an electron-hole symmetric resistance fitted to the hole values. } \label{capacitance}
\end{figure}

From the fits of the full set of admittance spectra we obtain the gate capacitance $C_g$ and the conductance $R_g^{-1}$ as function of gate voltage as displayed in Fig.\ref{capacitance}. The  capacitance and the conductance show a broad minimum at $V_{g}^{np}=0.67\;\mathrm{V}$ (sample E9-Zc) and $V_{g}^{np}=-0.07\;\mathrm{V}$ (sample C7-F) which we have identified as the charge neutrality point, shifted from zero by chemical doping.  The solid lines are  theoretical expectations for $C_g^{-1}(V_{g})$  with $C_{geo}=Const.$ and the finite temperature expression in Eq.(\ref{Cq}).
 Deviation from theory at low carrier density in  Fig.\ref{capacitance} is  likely due to inhomogeneity in the chemical doping which is not taken into account in Eq.(\ref{Cq}). Indeed it is more pronounced in sample E9-Zc which has a larger doping.  The fit at large density gives an accurate determination of the  geometrical capacitance  $c_{geo}=6.4\pm0.5\;\mathrm{fF\mu m^{-2}}$ (sample E9-Zc) and $c_{geo}=5.9\pm0.5\;\mathrm{fF\mu m^{-2}}$ (sample C7-F) in agreement with the rough estimates from geometry.  In the following $C_{geo}^{-1}$ is subtracted from $C_g^{-1}$ to obtain $C_Q$.  $R_g$  shows an excess at large electron density ($V_{g}\sim1\;\mathrm{V}$) which can be estimated from the difference of the measured resistance to the electron-hole symmetric expectation (dotted guide line in Fig.\ref{capacitance}). We assign this difference to the contact resistance due to the formation of a p-n junction in the access region with  $R_c/W\lesssim 300\;\mathrm{Ohms/\mu m}$. This value is consistent with theoretical expectation and experiment \cite{Huard2008PRB}. We also checked that the effect of $R_c$ is also seen in the corresponding admittance spectra (not shown) with an increase in the series resistance at large $V_{g}$. For this reason the electron regime is thereafter disregarded. The gate conductance $R_g^{-1}$ depicted in  Fig.\ref{capacitance} is similar to the drain source conductance measured in graphene transistors of similar size. The order of magnitude of the mobility ($\mu\simeq4500\;\mathrm{cm^2V^{-1}s^{-1}}$ at $n_c\simeq 10^{12}\;\mathrm{cm^{-2}}$) is also typical of room temperature behavior \cite{Chen2008Nnano}.

\begin{figure}
\includegraphics[scale=1]{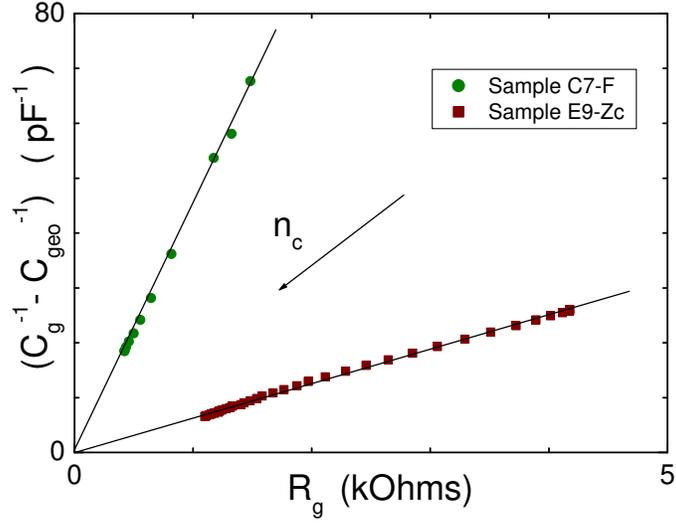}
\caption{Inverse gate capacitance as function of gate resistance. The geometrical contribution $1/C_{geo}$ is subtracted to isolate the quantum capacitance term. Data correspond to the hole conduction regime where contribution from the contact resistance can be neglected. The observed linear dependencies indicate an energy independent diffusion constant. The slopes $\omega^*$ give ${\cal D}=L^2/3\omega^*\simeq 180\;\mathrm{cm^2s^{-1}}$ (sample E9-Zc) and ${\cal D}\simeq 540\;\mathrm{cm^2s^{-1}}$ (sample C7-F).  } \label{scattering}
\end{figure}

Relying on the overall good quantitative agreement of the admittance spectra with the 1D rf field model and that of the gate voltage dependence of the capacitance, we proceed below to the quantitative analysis and deduce  in Fig.\ref{scattering} the diffusion constant ${\cal D}$ from the ratio $\omega^*=L^2/3{\cal D}$ of the measured quantum  capacitance $C_Q$ to the  charging resistance $R_{g}$.
In both cases we observe a linear dependence $\sigma(c_Q)$ which corresponds to an energy-independent diffusion constant ${\cal D}\simeq180\;\mathrm{cm^2/s}$ (sample E9-Zc) and $ {\cal D}\simeq540\;\mathrm{cm^2/s}$ (sample C7-F) and scattering  times $\tau(k_F)=2D/v_F^2$. This corresponds to scattering lengths $v_F\tau\simeq 40\;\mathrm{nm}$ and  $\simeq 100\;\mathrm{nm}$  in agreement with standard estimates \cite{Monteverde2010PRL}.

\begin{figure}
\centerline{\includegraphics[scale=0.7]{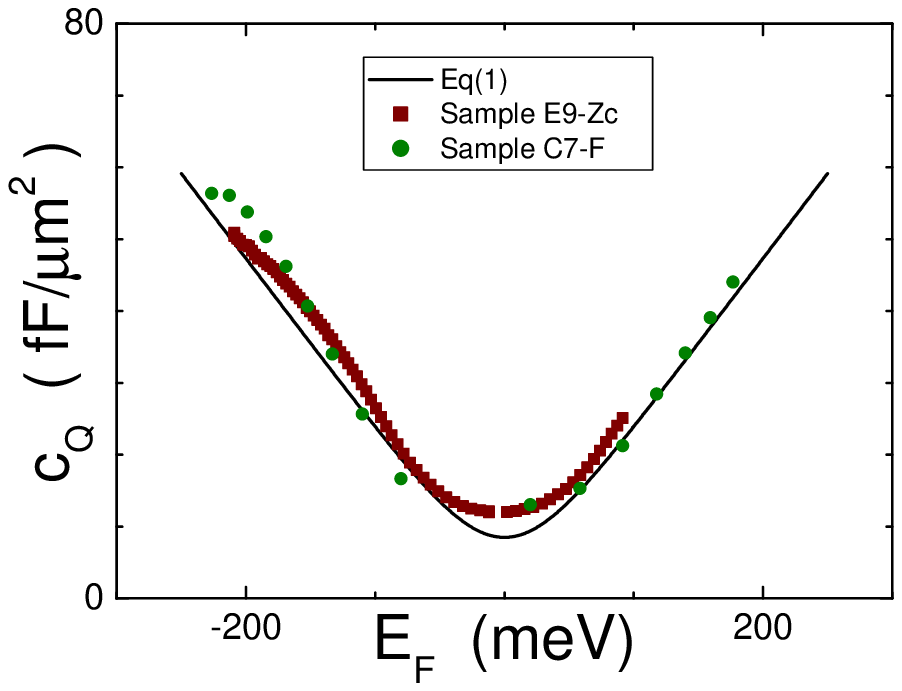}
\includegraphics[scale=0.7]{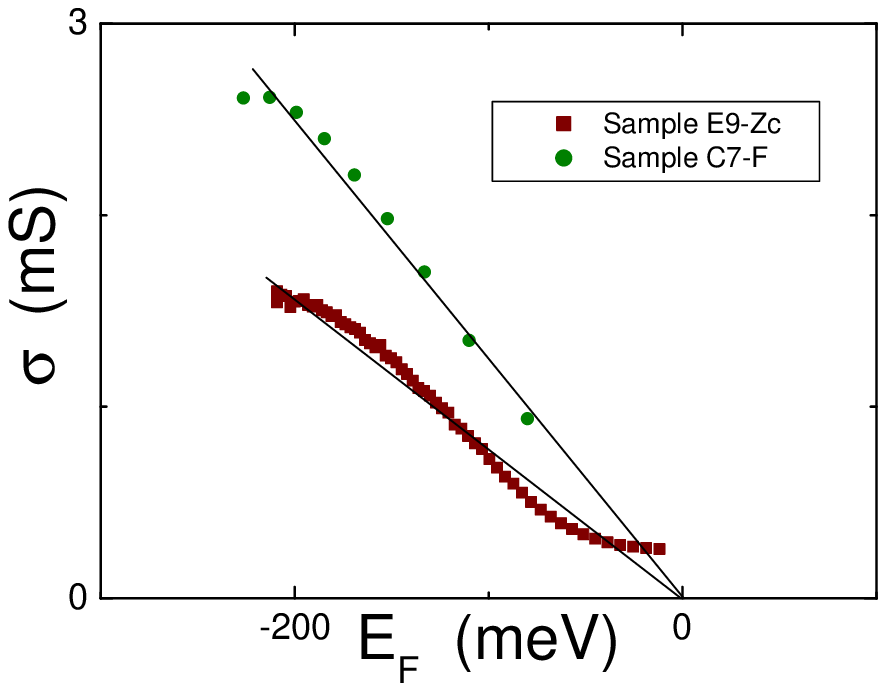}}
\caption{Chemical potential dependence of the quantum capacitance (a) and the conductivity (b) deduced form data in Fig.\ref{capacitance} using Eq.(\ref{E_F}).} \label{mu_plots}
\end{figure}

Using the experimental  $c_Q(V_{g})$ and  Eq.(\ref{E_F}) we can plot $c_Q(E_F)$ and $\sigma(E_F)$ in Fig.\ref{mu_plots}.  Small deviations from the theoretical estimate for $c_Q$ in Eq.(\ref{Cq}) are observed in both samples which are due to experimental uncertainties and disorder contribution. Uncertainties are yet too large for a quantitative estimate of the disorder contribution.  The precision of the high-frequency compressibility measurement can be increased by using larger flakes (more accurate deembedding), thinner gate oxide (smaller electrostatic gate impedance) and working at low temperatures (larger contrast in $c_Q$). The conductivity data merely follow the density of states reflecting the fact that  ${\cal D}=Const.$

\section{Discussion}

We have observed a strong sublinear dependence of conductivity in a broad carrier density range at room temperature. Direct comparison between conductivity and compressibility in the capacitor geometry at room temperature shows that this sublinear behavior extends to the low density limit and is well described by an energy-independent scattering time or equivalently $\sigma\propto\sqrt{n_c}$.  This result deviates from standard low temperature behavior ($\tau\sim k_F$) explained by  charged impurity,  resonant scattering models, or standard ripples. It could be accounted in a ripple scenario by taking a peculiar value for the correlation exponent,  $2H=3/2$, which deviates from standard expectation $H\simeq1$ \cite{Katsnelson2008PhilTrans,DasSarmaRMP2010}. Another possible mechanism, less discussed in the literature, would be Dirac-mass disorder proposed in Ref.\cite{Ziegler2006PRL} and characterized, according to a numerical calculation, by  $\sigma\propto E_F$ ($\tau=Const.$).

In conclusion, our RF admittance measurement of graphene capacitors provides new insight into the energy dependence of the transport scattering time.  We report on a new behavior in top gated graphene at room temperature characterized by a density independent scattering time.  Our experiment gives also access to the electronic compressibility via the quantum capacitance in fair agreement with theory.  It can be implemented at low temperatures using a cryogenic probe station and used to characterize the different scattering mechanisms in more detail. Our study may also prove useful for the design of high frequency graphene transistors.\vspace{1cm}

\begin{acknowledgments}
Authors warmly thank J.-N. Fuchs for fruitful discussions  drawing our attention to the Dirac-mass disorder model. We also acknowledge fruitful discussions with C. Glattli, P. Hakonen, H. Happy and G. Dambrine. The research has been supported by the contracts Cnano "GraFet-e", the ANR-05-NANO-010-01-NL-SBPC and ANR-xx-xx-MIGRAQUEL.
\end{acknowledgments}

 \newpage

\end{document}